\documentclass[twocolumn,prb,aps,floatfix]{revtex4}
\newcommand{\bc}{\begin{center}}
\newcommand{\ec}{\end{center}}

\usepackage[dvips]{color,graphics,epsfig,rotating}
\usepackage{graphicx}
\usepackage{bm}

\newcommand\lb         {\bm{l}} 
\newcommand\Ccal       {\mathcal{C}} 

\begin{document}
\preprint{}
\title{Kinetic Monte Carlo Simulations of Crystal Growth in
 Ferroelectric Alloys}

\author{Malliga Suewattana, Henry Krakauer, and Shiwei Zhang}
\affiliation
{Department of Physics, College of William and Mary\\
Williamsburg, VA 23187}

\date{\today}

\begin{abstract}
The growth rates and chemical ordering of ferroelectric alloys are studied with
kinetic Monte Carlo (KMC) simulations using an electrostatic model with long-range
Coulomb interactions, as a function of temperature, chemical
composition, and substrate orientation. 
Crystal growth is characterized by thermodynamic processes
involving adsorption and evaporation, with solid-on-solid
restrictions and excluding diffusion. A KMC algorithm is formulated to simulate
this model efficiently in the presence of long-range interactions.
Simulations were carried out on Ba(Mg$_{1/3}$Nb$_{2/3}$)O$_3$ 
(BMN) type materials. 
Compared to the simple rocksalt ordered structures, ordered BMN grows only at very low temperatures and
only under finely tuned conditions.
For materials with tetravalent compositions, such as 
$(1-x) {\rm Ba(Mg}_{1/3}{\rm Nb}_{2/3}{\rm )O}_3 
+ x {\rm BaZrO}_3$ (BMN-BZ), the model does not 
incorporate tetravalent ions at low-temperature, exhibiting a phase-separated ground state instead. At higher temperatures,
tetravalent ions can be incorporated, but the resulting crystals show no chemical ordering in
the absence of diffusive mechanisms.
\end{abstract}

\maketitle

\section{Introduction}

Ferroelectric crystals are known for their important technological
applications such as high-permitivity dielectrics, piezoelectric
sensors, transducers, and mechanical actuators. \cite{uchino96}  Recently,
single-crystal relaxor perovskites such as
Pb(Zn$_{1/3}$Nb$_{2/3}$)O$_3$-PbTiO$_3$ (PZN-PT) and
Pb(Mg$_{1/3}$Nb$_{2/3}$)O$_3$-PbTiO$_3$ (PMN-PT) were synthesized and
found to exhibit ultrahigh stain and very large piezoelectric
constants. \cite{ParkandShrout}  The structure of alloys like PMN-PT
can be viewed as a perovskite ABO$_3$ framework (a cubic lattice for
the ideal perovskite crystal), with Pb ions on the A-site and a solid
solution of (Mg$^{+2}$, Nb$^{+5}$, Ti$^{+4}$) ions on the B-sites,
with average $+4$ B-site ionic charge. Of course this is an idealized
picture, neglecting vacancies, impurities, local structural
distortions, and partial chemical ordering on the B-sites.  

Partial B-site chemical ordering is a common feature of 
the high-piezoelectric solid solutions. While random B-site ordering is observed in 
isoelectronic solid solutions like Pb(Zr$_{1-x}$Ti$_x$)O$_3$ (PZT), 
non-isoelectronic B-site solid solutions A(BB$^{'}$B$^{''}$)O$_3$,
with  B-site cations from group II, IV, and V,
often exhibit compositionally-dependent B-site chemical ordering.
At 1640$^{\circ}$C, when the tetravalent composition $x$ is increased in
$(1-x) {\rm Ba(Mg}_{1/3}{\rm Nb}_{2/3}{\rm )O}_3$
~+~$x$~BaZrO$_{3}$ (BMN-BZ), the
following sequence of B-site ordering is observed: [111]$_{1:2}$ order
for $x<5\%$; then [111]$_{1:1}$ order for $5\%<x<25\%$; and finally
disorder for larger $x$. \cite{ad} 
The [111]$_{1:2}$ notation refers to x-ray observation of alternating 
$\beta \beta \beta'$ [111] stacking of B-sites, where
$\beta$ and $\beta'$ denote average scattering sites. For example,
in BMN-BZ with $x=0$, one can identify $\beta$ with Nb and $\beta'$ with Mg.
The [111]$_{1:1}$ notation refers to x-ray observation of rocksalt-like 
alternating $\beta \beta'$ [111] stacking of B-cations. In this case,
the assignment of the $\beta$ and $\beta'$ sites has been debated, as
discussed below in connection with the space-charge and 
random-site models \cite{ad}.
Other Ba-based perovskites, e.g.,
(1-$x$)~Ba(Mg$_{1/3}$Ta$_{2/3}$)O$_{3}$~+~$x$~BaZrO$_{3}$ (BMT-BZ),
\cite{ad} (1-$x$)~Ba(Mg$_{1/3}$Nb$_{2/3}$)O$_{3}$~+~$x$~BaZrO$_{3}$
(BMN-BZ), \cite{cadp} display a similar sequence of B-site order.  On
the other hand, for Pb-based systems, e.g.,
(1-$x$)~Pb(Mg$_{1/3}$Ta$_{2/3}$)O$_{3}$~+~$x$~PbZrO$_{3}$ (PMT-PZ),
[111]$_{1:2}$ order is not observed at $x$~=~0; instead, annealing
between 1325$^{\circ}$C and 1350$^{\circ}$C results in [111]$_{1:1}$
order all the way down to $x$~=~0. \cite{ad2,ad3} Other Pb-based
perovskites , e.g., Pb(Mg$_{1/3}$Nb$_{2/3}$)O$_3$ (PMN),
\cite{Husson,Chen} display similar B-site ordering.

Since
their discovery, growing large single crystals has been 
a major research goal, but this effort has 
been largely unsupported by theory, because of the difficulty in
modelling and simulating the non-equilibrium processes
occurring in nucleation and crystal growth in such complex materials. 
In this paper, we use kinetic Monte Carlo 
\cite{kmc-bortz-75} simulations of a simple effective Hamiltonian
to model the growth process of these ferroelectric crystals. 

Given the ionic character of these materials, it is not surprising that
the inclusion of Coulomb interactions has been found
to be crucial in describing their properties. 
A simple, purely electrostatic model introduced
by Bellaiche and Vanderbilt (BV) \cite{Bellaiche-Vander-98}
has had considerable success in explaining
the observed equilibrium B-site chemical ordering in many 
perovskite alloys. 
The BV model only considers
Coulomb interactions between point charges 
(+2, +5, +4, etc., representing the different atomic species)
that reside on the B-sites, which are
constrained to lie on an ideal cubic sublattice.

This electrostatic model is the starting point of our growth simulations.
Simplified models
based on Ising like effective Hamiltonians $H_{\rm eff}$ 
have been used to model growth in
simpler systems \cite{Jackson95,Shneidman99}.  
These models often have only short-range
interactions. 
To adapt the electrostatic model of BV to study crystal growth, 
we consider a slab-geometry with periodic
boundary conditions in two-dimensions only.  The slab is viewed as
being embedded in a liquid-phase melt, which is parametrized by a
chemical potential difference with the solid bulk phase. 
In our simulations, a solid-on-solid (SOS) restriction is imposed, which 
requires that adsorption only occur onto empty lattice sites
directly above an occupied site, so void formation is neglected. In
keeping with the simplicity of the model, diffusion in the bulk and at
the surface is also neglected.

The non-equilibrium dynamics of the growth process is modeled using the 
kinetic Monte Carlo (KMC) method \cite{Binder2}.  The KMC algorithm
introduced by Bortz, Kalos, and Lebowitz (BKL) \cite{kmc-bortz-75} has
been quite successful in simulating crystal growth in Ising-like
models with short-range interactions between adatoms. 
We generalize the algorithm to efficiently handle 
the long-range interactions
in the BV model. 

In this paper we study the properties of this minimal paradigm of the
growth process to determine if it can yield insights into the physics of the 
observed B-site chemical ordering. Section II presents our
theoretical approach. The model of BV 
is reviewed, and our adaptation of it for the growth modeling is 
discussed, including the special handling of electrostatic
interactions during the growth process. Our generalization 
and modification of the KMC algorithm are then described which 
allows efficient treatment of the  long-range Coulomb interactions.
Section III presents the
results of our growth simulations for A(BB$^{'}$)O$_3$ and 
A(BB$^{'}$B$^{''}$)O$_3$ crystals.
To help understand our growth results for the latter systems, 
where a (typically 
small) fraction of tetravalent B$^{''}$ ions are mixed in, we also carry out 
total energy calculations to study their stability. Finally
in section IV we further discuss our results and prospects for the model.
In the appendix, we include some technical details on the treatment of 
the long-range 
interactions in our simulations. 
A preliminary account of part of 
this work has already appeared \cite{Tahan2001}.

\section{Theoretical Approach}
At each stage of the simulation the crystal is modeled as a slab of finite
thickness. However, it is convenient to index the allowed B-sites
as in an infinite three-dimensional crystal lattice
\begin{equation}
\label{indexsites}
\bm{l}  = i\, \bm{a}_1 + j\, \bm{a}_2 + k\, \bm{a}_3 \, .
\end{equation}
Two-dimensional (2-D) periodic boundary conditions (PBC) are employed along
the $\bm{a}_1$ and $\bm{a}_2$ directions, which define the $x$-$y$
Cartesian plane, using a $L_1 \bm{a}_1$~$\times$~$ L_2 \bm{a}_2$~$=$~$\bm{A}_1$~$\times$~$\bm{A}_2$ 2-D
supercell.  Growth proceeds along the $z$-direction.  The simulation
is initialized as a slab of uniform thickness $ H_0 \bm{a}_3$, with a
predefined B-atom configuration.  
A given simulation is terminated when either the maximum slab thickness
or the maximum number of Monte Carlo (MC) time steps is reached. We 
use the notation $L \times L \times H_{\rm max}$ to label
a particular simulation, where $H_{\rm max}$ is the number of layers 
for maximum slab thickness 
(the initial substrate included). 

The SOS restriction that we impose does not allow the formation of voids.
The crystal
configuration, $\Ccal$, is specified at each stage of the simulation 
by the set of occupied sites $\lb = (i,j,k)$ and their charges
$q_{\lb}$.  The BV electrostatic model cannot be directly used in this
slab geometry, due to ill-defined electrical boundary conditions in
the $z$ direction and the lack of exact charge neutrality during the
growth simulation. Section A describes how we handle these issues.
Similarly, a direct application of the KMC algorithm is 
inefficient due to the long-range Coulomb interaction.  Section B
describes the KMC method and our modifications to make 
it applicable to the model.

\subsection{The Electrostatic Model}
\label{sec:BVmodel}

The BV model is derived by considering the 
total electrostatic energy 
for an A(BB$^\prime$B$^{''}$)O$_3$ compound:
\begin{equation}
\label{BV}
E(\Ccal) = \sum_{(\lb\tau,\lb^\prime\tau^\prime)}  
{Q_{\lb\tau} Q_{\lb^\prime\tau^\prime} \over 
\epsilon \big|{\bm R}_{\lb\tau} - {\bm R}_{\lb^\prime\tau^\prime}\big| },
\end{equation}
where ${\bm R}_{\lb\tau}$ is the position of the ion on site $\tau$
($=$A, B, O$_1$, O$_2$, O$_3$)
in cell $\lb$, and $\epsilon$ is the
dielectric constant. 
For a given Bravais lattice,
$\epsilon$ sets the energy scale.
We consider the perovskite structure with 
group II A-site atoms (e.g. Ba, Pb), so the charges on
the A and O sites have
fixed values of $+2 e$ and $-2 e$, respectively. Since the average B-site charge
is $+4 e$, it is convenient to express the charges
on the B-sites, $Q_{\lb,B}$, as
\begin{equation}
\label{q_lb}
Q_{\lb,B} = 4 e + q_{\lb}.
\end{equation}
Up to a constant, the configurationally averaged electrostatic energy
depends only on the B-site charges, since the
configurational average of $q_{\lb}$ is zero:
\begin{equation}
\label{BV_Bsite}
E_B(\Ccal) = {1\over \epsilon a} \sum_{(\lb,\lb^\prime)}
{q_{\lb} q_{\lb^\prime} \over
 \big|{\bm l} - {\bm l^\prime}\big| },
\end{equation}
where we have for simplicity restricted ourselves in Eq.~(\ref{BV_Bsite}) to a cubic Bravais lattice with 
lattice parameter $a$, and  ${\bm R}_{\lb B} = {\bm l}\, a$.
In this model each cell $\lb$ is therefore reduced to a single lattice
site with charge $q_{\bm{l}}$, and the energy of the compound is
given by the inter-site Coulomb interaction. 

The long-range Coulomb interaction must be treated with 
care in a bulk simulation to ensure proper convergence. 
For 2-D and 3-D simulations with periodic boundary conditions, 
the method of Ewald  
is often used, in which periodic images of the charges and neutralizing
background charges are introduced 
\cite{deLeeuw80,deLeeuw75,Nijober,Ceperley,Smith83} 
so that the bare Coulomb form 
$1/|{\bm l} - {\bm l^\prime}|$ is replaced with a reduced form
$v({\bm l} - {\bm l^\prime})$. For our growth simulations, we are dealing 
with a slab geometry with PBC only 
in two dimensions ($x$-$y$).
Some modifications
are required before the Ewald method can be applied.

In the simulations, we will need to calculate the energy change 
from Eq.~(\ref{BV_Bsite}) due to
the evaporation of a charged ion $q_{\bm{l}'}$ at the surface of
the crystal (see
Eq.~(\ref{eq:V_ewald}) below).
The distribution of point charges that $q_{\bm{l}'}$ ``sees''
can be described by the charge density
\begin{equation}
\rho ({\bm{r}}) = \sum\limits_{\bm{l }} {\sum\limits_{\bm{R}} {q_{\bm{l }} 
 {\delta ({\bm{r - l  - R}})}} } 
\end{equation}
where $\bm{l}$ runs through the position vectors of the atoms within the
simulation cell, and $\bm{R}$ is a 2-D Bravais supercell lattice vector: ${\bm{R}}
= n_1 {\bm{A}}_1 + n_2 {\bm{A}}_2 $.
Directly summing the Coulomb potentials of the individual point
charges, $V({\bm{r - l - R}}) = q_{\bm{l }} /|{\bm{r - l -
R}}|$, leads to an ill-defined and conditionally convergent result, as
is well known. However, for three-dimensional periodic boundary
conditions, a unique solution of Poisson's equation exists (for an
electrically neutral system), and it is conveniently calculated using
Ewald's method. 
Subject to some
additional, physically motivated conditions, a unique solution can
also be found for finite thickness slabs that are infinite in extent
along two spatial directions.

Solutions of Poisson's equation, 
$\nabla^2\/V({\bm{r}})\/=\/-\/4\pi\/\rho({\bm{r}})$, in our simulations are
subject to two-dimensional (2-D) PBC
$V({\bm{r}} + {\bm{R}}) = V({\bm{r}})$, as is the charge density
$\rho ({\bm{r}})$.
The 2-D PBC
imply that $V({\bm{r}})$ and $\rho({\bm{r}})$ can be expanded as:
\begin{equation}
\begin{array}{l}
 \rho ({\bm{r}}) = \sum\limits_{\bm{G}} {\rho _{\bm{G}} (z)e^{i{\bm{G}}\cdot
{\bm{r}}_p } }  \\ 
 V({\bm{r}}) = \sum\limits_{\bm{G}} {V_{\bm{G}} (z)e^{i{\bm{G}}\cdot
{\bm{r}}_p } } , \\ 
\end{array}
\label{expand_expGr}
\end{equation} 
where $\bm{G}$ is a 2-D supercell reciprocal lattice vector and $\bm{r}_p$
is the $x$-$y$ component of $\bm{r}$, 
$\bm{r}_p =  \bm{r} - (\bm{r} \cdot \bm{\hat z})\bm{\hat z} =   
i {\bm{a}}_1  + j {\bm{a}}_2 $.
Substitution of Eqs.~(\ref{expand_expGr}) into Poisson's equation 
yields the ordinary differential equation
\begin{equation}
\frac{{d^2 V_{\bm{G}} (z)}}{{dz^2 }} - G^2 V_{\bm{G}} (z) =  - 4\pi \rho _{\bm{G}} (z),
\label{Poisson_Gz}
\end{equation}
whose solution can be expressed as
\begin{equation}
V_{\bm{G}} (z) =  - 4\pi \int\limits_{ - \infty }^\infty  {\mathcal{G}(z - z')\rho _{\bm{G}} (z')dz'} ,
\label{V_quadr_G}
\end{equation}
where $\mathcal{G}(z-z')$ is the Green's function corresponding to 
Eq.~(\ref{Poisson_Gz}).

If there are any ill-defined contributions to the Coulomb potential,
they must arise from the $\bm{G} = \bm{0}$ solution in Eq.~(\ref{Poisson_Gz})
or (\ref{V_quadr_G}). This is because only the $\bm{G} = \bm{0}$ term of
$\rho({\bm{r}})$ in Eqs.~(\ref{expand_expGr}) contributes to the net
slab charge. Even if the slab is electrically neutral, there may still
be a net dipole moment $D$, which would lead to different asymptotic values
of Coulomb potential at $z = \pm \infty$. Again, $D$
also depends only on the $\bm{G} = \bm{0}$
term of $\rho({\bm{r}})$, where
\begin{equation}
D \equiv \int\limits_{ - \infty }^\infty  {z\bar \rho (z)dz},
\end{equation}
with
\begin{equation}
\bar \rho (z) = \frac{1}{A}\int_A {\rho ({\bm{r}})} dxdy = 
\rho _{{\bm{G}} = \bm{0} } (z),
\end{equation}
where $A$ is the area of the 2-D supercell.

We therefore first consider the well-defined $\bm{G} \ne \bm{0}$ solutions of Eq.~(\ref{Poisson_Gz}).
Physically meaningful results require that the
solutions satisfy 
$\mathop {\lim }_{\left| z \right| \to \infty} V_{\bm{G}} (z) = 0$, 
which leads to the following unique definition
of the $\bm{G} \ne \bm{0}$ Green's function:
\small
\begin{equation}
\mathcal{G}(z \!- \!z') \equiv  
- \frac
{\left[ {\vartheta (z\!-\!z')e^{ -\! G(z\!-\!z')}  
\!  +\! \vartheta (z'\!-\!z)e^{G(z\!-\!z')} } \right]}
{2G},
\label{Green_G}
\end{equation}
\normalsize
where $G=|\bm{G}|$.
For any reasonably localized charge distribution $\rho _{\bm{G}}(z)$,
Eqs.~(\ref{V_quadr_G}) and (\ref{Green_G}) result in well-behaved,
exponentially decaying solutions of $V_{\bm{G}}(z)$ as $\left| z \right|
\to \infty $.

For $\bm{G} = \bm{0}$, Eq.~(\ref{Poisson_Gz}) becomes 
\begin{equation}
\frac{{d^2 V_{\bm 0} (z)}}{{dz^2 }} =  - 4\pi \rho_{\bm 0} (z).
\label{Poisson_G0z}
\end{equation}
As adatoms are adsorbed or atoms evaporate in the course of the growth
simulations, the net charge will fluctuate so that the total charge in
the simulation supercell will not be precisely zero at each stage of
the simulation. Similarly a net dipole $D$ may form.  However,
in a real growth process there are always compensating charges that
will cancel any ill-defined long-range effects due to the lack of
charge neutrality or the presence of a dipole moment. In our
calculations, we simulate this by a construction that ensures that
$\rho _0 (z)$ in Eq.~(\ref{Poisson_G0z}) always represents a neutral
charge distribution with $D = 0$. This leads to well-defined
boundary conditions 
$\mathop {\lim }_{\left| z \right| \to \infty} V_{\bm{0}} (z) = 0$.

As in the 3-D Ewald method, a diffuse localized charge density
$g(\bm{r})$ is added and subtracted to each point charge to facilitate
the decomposition of the potential into absolutely convergent
direct- and reciprocal lattice sums:
\begin{widetext}
%
\begin{eqnarray}
 \rho ({\bm{r}}) & = & 
\sum\limits_{\bm{l }} {\sum\limits_{\bm{R}} {q_{\bm{l}} \left[ {\delta ({\bm{r - l  - R}}) - g({\bm{r - l  - R}})} \right]} }  
+\sum\limits_{\bm{l }} {\sum\limits_{\bm{R}} {q_{\bm{l}} g({\bm{r - l  - R}})} } \nonumber  \\ 
  &\equiv & \rho _1 ({\bm{r}}) + \rho _2 ({\bm{r}})\,. 
\label{rho1rho2}
\end{eqnarray}
%
\end{widetext}
The diffuse charge density $g(\bm{r})$ is chosen to be a normalized
spherically symmetric Gaussian, as in the 3-D Ewald method:
\begin{equation}
g({\bm{r}}) \equiv \left( {\frac{\alpha }{\pi }} \right)^{3/2} e^{ - \alpha r^2 },
\end{equation}
where the value of the Ewald convergence parameter $\alpha$ is
arbitrary, but is usually chosen to optimize the convergence of both
the direct- and reciprocal-lattice sums. The integrated charge of
$\rho_1({\bm r})$ is zero by construction, as is its dipole moment $D$,
so its contribution $V_1(\bm{r})$ to the Coulomb potential can be
obtained by a rapidly convergent direct-lattice sum, given in the Appendix.

The procedure for calculating the Coulomb potential $V_2(\bm{r})$ due
to $\rho_2 (\bm{r})$ requires special handling. 
$ V_2 ({\bm{l'}})$, the potential at the position of $q_{\bm{l' }}$ in the
simulation cell, is due to: {\em i\/}) the $\bm{l} \ne \bm{l'}$ Gaussian
charge densities and their periodic images $q_{\bm{l }}
g(\bm{r-l-R})$, and {\em ii\/}) the periodic images $q_{\bm{l' }}
g(\bm{r-l'-R})$. [As in the 3D Ewald method, a spurious interaction
of the point charge $q_{\bm{l' }}$ with its own Gaussian density
$q_{\bm{l' }} g(\bm{r-l'})$ is explicitly removed
later.] Alternatively, the contribution ({\em ii\/}) above can be replaced by
{\em iia\/}) 
the Gaussian densities $-q_{\bm{l }} g(\bm{r-l'-R})$ located at the 
positions of the periodic images of $\bm{l'}$.
In a bulk crystal
simulation with 3-D PBC and a neutral simulation cell, 
these two formulations are equivalent, since
the
integrated total charge 
vanishes:
\begin{equation}
\sum\limits_{{\bm{l }} \ne {\bm{l }}'} 
{-q_{\bm{l }} }  =  q_{{\bm{l }}'}\,.
\end{equation}

In the 2-D slab geometry of our
growth simulations, this will not be the case in general.
Overall charge neutrality is still satisfied in a statistical sense, however.
Our procedure for calculating $V_2(\bm{r})$ consists of two approximations. 
The first approximation is to use 
formulation ({\em iia\/}) above. 
This means that the contribution of each
$q_{\bm{l }} g(\bm{r})$ sublattice to $ V_2 ({\bm{l'}})$ is to be
calculated as the potential due to the charge density:
\begin{equation}
\label{rhosubl_a}
\rho_2 ^{({\bm{l }},{\bm{l }}')} ({\bm{r}}) = q_{\bm{l }} \sum\limits_R 
{\left[ {g({\bm{r - l  - R}}) - g({\bm{r - l }}'{\bm{ - R}})} \right]}.
\end{equation}
Since the integrated charge of $\rho_2 ^{({\bm{l }},{\bm{l }}')}
({\bm{r}})$ is zero, the use of this approximation effectively
imposes overall charge neutrality at each stage of the growth simulation.

The boundary conditions are still ill-defined however, since the sum
of sublattice potentials due to the $\rho_2 ^{({\bm{l }},{\bm{l
}}')} ({\bm{r}})$ may still have a dipole moment $D$. We therefore
introduce a second approximation: the Gaussian image densities
$-q_{\bm{l }} g(\bm{r-l'-R})$ are made coplanar with the $q_{\bm{l }}
g(\bm{r-l-R})$ sublattice. In other words, the Gaussian densities
$-q_{\bm{l }} g(\bm{r})$ are placed at positions that are the
projections of the $q_{\bm{l' }}$ image positions onto the
plane defined by the $q_{\bm{l }}$ sublattice. 
In place of Eq.~(\ref{rhosubl_a}), the contribution of
each $q_{\bm{l }} g(\bm{r})$ sublattice is thus calculated as the
potential due to the charge density:
\begin{equation}
\tilde \rho_2 ^{({\bm{l }},{\bm{ l }}')} ({\bm{r}}) = q_{\bm{l }} \sum\limits_R 
{\left[ {g({\bm{r - l  - R}}) - g({\bm{r - \tilde l }}\/'{\bm{ - R}})} \right]},
\label{rhotilde}
\end{equation}
where $\bm{ \tilde l' }$ denotes the projection of the position
$\bm{l' }$ onto the plane defined by the $q_{\bm{l }}$ sublattice. 
The charge density $\tilde \rho_2 ^{({\bm{l }},{\bm{ l }}')} ({\bm{r}})$ 
has a rapidly convergent
expansion in terms of 2-D planewaves given by Eq. (\ref{expand_expGr}).
Moreover, the $\bm{G}$~=~0 contribution of 
$\tilde \rho_2 ^{({\bm{l }},{\bm{ l }}')} ({\bm{r}}) $ vanishes, so 
the Coulomb potential $V_2(\bm{r})$ is readily found using 
Eqs.~(\ref{V_quadr_G}) and (\ref{Green_G}). 
These two approximations ensure overall average-charge neutrality and
vanishing dipole moment $D = 0$, resulting in a well-defined Coulomb
potential at each stage of the growth simulation. Complete formulas
for the potential $v ({\bm{l }}'{\bm{ - l }})$ are given in the Appendix.
 
\subsection{Kinetic Monte Carlo method for long-range interactions}
\label{sec:KMCnew}

The kinetic Monte Carlo (KMC) method is one of several simulation
techniques commonly employed to model the relaxation processes 
of systems away
from equilibrium (e.g. growth processes). It has been applied
successfully to crystal growth and surface/interface 
phenomena, \cite{Binder2,LeviandKotrla} mostly in the context of kinetic Ising
models.
Due to the long-range interactions between ions in our electrostatic model,
the usual implementation of KMC for Ising-like models is inefficient, 
with the acceptance
rates of events becoming very low. We developed a modified sampling 
algorithm to make the simulation practical for this model. 
Here we briefly outline the basic theoretical background for the KMC
method, and then describe our modifications and give the relevant 
implementation details.
 
In the KMC simulation, the dynamics of the system is described as 
stochastic processes such
as adsorption, evaporation, and surface migration. We consider only 
the first two in our simulation. 
As mentioned, the adatoms represent the B-site ions in the single crystal
perovskite alloy. They are characterized entirely by their charges
and they interact with each other by the interaction described above.

In the grand canonical ensemble, 
the Hamiltonian that will be used in the growth simulations can then
be expressed in term of Eq.~(\ref{BV_Bsite}) as
\begin{equation}
\label{hamiltonian}
{\cal{H}}(\Ccal) = E_B(\Ccal) + \Delta \mu N,
\end{equation}
where $N$ is the total number of adsorbed adatoms. 
The electrostatic energy term in the Hamiltonian is responsible for evaporation,
while the second term, which depends on the chemical potential difference
between the solid and the gas phases, controls the
 rate in which adatoms stick on the surface.
 The growth simulation is then characterized by competing
adsorption  and desorption events.
The SOS
restriction imposed in the simulation
prevents formation of vacancies and 
allows us to write $H$ as
\begin{equation}
\label{hamiltonian2}
{\cal{H}}(\Ccal) = E_B(\Ccal) + \Delta \mu 
\sum_{i,j} h_{ij},
\end{equation}
where $h_{ij}$ is the number of layers in the present crystal configuration 
at the horizontal position $i \bm{a}_1$~$+$~$ j \bm{a}_2$.

In KMC the time evolution of the
system is simulated through a Markov chain of configurations.  
Let us define $P(C,t)$
as a time-dependent distribution of configurations.
The transition rate from $C$ to
$C^\prime$, a crystal configuration
related to $C$ by a single time step, 
is denoted by $w(C \rightarrow C^\prime)$. 
We then have the usual master equation \cite{LeviandKotrla}:
\begin{eqnarray}
\frac{\partial P(C,t)}{\partial t} = 
&-&\sum_{C'}w(C \rightarrow C') P(C,t) \nonumber \\
&+&\sum_{C'}w( C' \rightarrow  C) P( C',t),
\label{eq-master}
\end{eqnarray}
where the first term on the right describes the loss because of
transitions away from $C$, while the second term describes the gain
because of transitions into $C$.  In the equilibrium limit (as $t
\rightarrow \infty$), the Boltzmann distribution
\begin{equation}\label{boltz-eq}
P_{eq} = Z^{-1} \exp\left[ \frac{-{\cal H}(\Ccal)}{k_B T} \right]
\end{equation}
is reached, where $k_B$ is the Boltzmann constant. We require that detailed balance be satisfied:
\begin{equation}
\frac{w(C \rightarrow C')}{w( C' \rightarrow
 C)} = \frac{P_{eq}(C')}{P_{eq}( C)} = \exp \left[
\!-\!\frac{{\cal H}( C') - {\cal H}( C)}{k_B T} \right].
\label{eq:DB}
\end{equation}

The KMC technique can be viewed as a method of solving Eq.~(\ref{eq-master}) 
stochastically. We adopt the following choice
of transition rates $w(C \rightarrow C')$
\begin{eqnarray}
w_{\rm a} &=& \exp{(\Delta\mu/k_B T)} 
\label{eq:rateabs}
\\
w_{\rm e} &=& \exp{(-\Delta E_B(\Ccal)/k_B T)}, 
\label{eq:rateevap}
\end{eqnarray}
where $w_a$ and $w_e$ are the rates for adsorption and evaporation,
respectively, of an adatom. It can be verified straightforwardly that 
this choice indeed satisfies Eq.~(\ref{eq:DB}). 
The rate $w_e$ for an adatom of charge $q_{\bm{\tau}'}$  to evaporate
from the surface
depends on the change in total
potential energy in the crystal
\begin{eqnarray}
\Delta E_B(C) &=&  E_B(C^\prime) - E_B(C) \nonumber\\
&=& 
{q_{\bm{l}'} \over \epsilon a}
\sum_{\bm{l} } q_{\bm{l}} v (\bm{l}'-\bm{l}).
\label{eq:V_ewald}
\end{eqnarray}

For kinetic Ising models, the algorithm of BKL\cite{kmc-bortz-75} allows
an efficient stochastic realization of the kinetic process under the
choice in Eq.'s~(\ref{eq:rateabs}) and (\ref{eq:rateevap}). 
In this algorithm, a site $(i,j)$ is
selected randomly in each step at the surface of the grown crystal. 
An event is then selected 
by Monte Carlo sampling \cite{Kalos-book} from the list of three
possible events, $\{{\rm adsorption},\ {\rm evaporation},\ {\rm nothing}\}$. 
The interaction in Ising type models is limited to near-neighbors, and
the energy difference $\Delta E_B(C)$ is completely determined by the
{\em local\/} environment at site $(i,j)$.  
The global maximum of $w_e$, i.e., 
the minimum possible energy change, 
$\Delta E^{\rm min} = \min[\Delta E_B(C)]$, 
can be obtained straightforwardly
by considering all possible local configurations. This gives a 
corresponding global maximum of the evaporation rates: 
$w_e^{\rm max} = \exp{(-\Delta E^{\rm min}/k_BT)}$, which defines
a normalization factor: 
\begin{equation}
W\equiv w_a + w_e^{\rm max}.
\label{eq:rates_Ising_norm}
\end{equation}
The relative probabilities for the three events are therefore
\begin{equation}
\{P_a\equiv \frac{w_a}{W}, P_e\equiv\frac{w_e}{W}, 
P_n\equiv1-P_a-P_e\}. 
\label{eq:rates_Ising}
\end{equation}



With the electrostatic model, however,  
the energy change 
in Eq.~(\ref{eq:rateevap}) depends on the {\em entire\/} configuration $C$.
It is therefore difficult to determine the global
minimum, $\Delta E^{\rm min}$. Indeed, even if 
$\Delta E^{\rm min}$ could be identified, the energy change
$\Delta E_B(C)$, which can vary greatly with $C$ and the simulation 
cell size,
would be much greater than
$\Delta E^{\rm min}$ for most configurations. This 
would cause the evaporation and adsorption probabilities 
$P_a$ and $P_e$ to be small, with 
$P_n$ approaching unity. As a result the acceptance rate of events 
becomes small, and the algorithm becomes ineffective.

To overcome this difficulty, we modify the standard algorithm 
so that all $N=L_1 \times L_2$ surface sites 
are considered {\em simultaneously\/}, instead of sweeping through
the surface sites. An event list is created 
which includes every
possible event for every possible surface site. 
This increases the algorithm complexity, because of the 
need to store and update an array
of surface potentials, calculate the event list, 
and sample an event from this list. 
The advantage is that an event is guaranteed to take place in each
step of the algorithm and that the need for determining $\Delta E^{\rm min}$
is eliminated.  Evaporation/adsorption rates for all possible
sites are normalized. The sum of the probabilities for an adsorption
or evaporation to occur at a surface site is unity. 
Specifically, 
the modified algorithm consists of the following steps:

\newcounter{j}
\begin{list}{(\roman{j})}
{\usecounter{j} \renewcommand{\baselinestretch}{1}\normalsize
\sffamily}
\item Generate a list, $\cal E$, of all possible events per time step.
There are $2N$ possible events: an evaporation or an adsorption could
happen on each of the $N=L_1 \times L_2$ surface sites.
\item Calculate the rates ($w$) of adsorption and evaporation for each
site on the surface.
Denote the total rates by $W$: 
$W=\sum\limits_i^{2N} w_i$.
\item Normalize these $2N$ rates by $W$, giving probabilities, $P_i$,
for adsorption and evaporation on sites $1, 2, \cdots, 2N$. 
\item Generate a random number $r \in [0,1)$ and choose the first
event ${\cal E}_i$ such that $\sum\limits_{k=1}^{i} P_k \ge r$. An event will
always be chosen.
\item Generate the new configuration $C'$ based on the chosen event
${\cal E}_i$.
\item Assign a ``real time'' increment 
$\Delta t_{\rm real}=-\ln(r^\prime)/W$ to this 
MC step, where $r^\prime$ is another random number on $[0,1)$.
\end{list}

The last step 
is a result of our considering the global 
event list and forcing an event to occur in every step. 
The issue 
of real ``time'' in a KMC simulation is a subtle one.  Often the
Monte Carlo time, $t_{\rm MC}$, is used as some measure of the real
time. In the standard algorithm, the global normalization factor
$W$ (defined by $w_e^{\rm max}\,$) controls the overall rate of events and
sets a ``time scale.'' In our approach, $W$ is time-dependent, and
an event is forced to happen 
in each step
regardless of the total rate $W$ for the configuration at hand.
When $W$ is low, an evaporation or adsorption is less likely to happen
but one is selected anyway. 
Conversely, when $W$ is high, an evaporation or adsorption is more likely 
to happen
but still only one is selected. This introduces a bias which should 
vanish in the limit of large system size but which should be
corrected for at finite $L$. Based on the rate equation, we assume 
an exponential relation between time and $W$. A step in which $W$ is high
corresponds to a short time, and vice versa. Step (vi) is a way to 
account for this time scale stochastically, by rescaling $\Delta t_{\rm MC}$ 
with a MC sampling from an exponential 
distribution which is determined by the normalization factor $W$ in each step.

\section{Results}

We now present the results from our simulations for A(BB$^{'}$)O$_3$ and
A(BB$^{'}$B$^{''}$)O$_3$ crystals.
Growth simulations are presented in Section A.  
Growth rates are studied, and charge-charge
correlation functions are
calculated to measure the degree of growth order.
The effects of varying
the crystallographic orientation of the slabs were explored,
with the slabs labelled according
to the slab perpendicular ($z$) direction. 
In A(BB$^{'}$B$^{''}$)O$_3$ systems, a fraction of tetravalent B$^{''}$ ions 
are mixed in. In our growth simulations, these tetravalent ions do not appear to 
mix at low temperatures, choosing instead to phase-separate from the pure crystal. 
To further study this, Section B presents results of static total energy and free-energy
calculations for fixed slab configurations.

\subsection{Crystal Growth}

The growth process 
is a function of temperature $T$, chemical potential difference $\Delta \mu$,
and the Coulomb interaction. These parameters are fixed
throughout a given simulation. 
As discussed in Section \ref{sec:BVmodel} and in the Appendix, we
tabulate $v (\bm{l}'-\bm{l})$, and we will use reduced units in our 
simulations below. The energies ($\,\Delta \mu$ and $E_B(C)\,$) are 
scaled by $\xi\equiv 1/\epsilon a$. There is only one free parameter 
between $\xi$ and the temperature 
$k_B T$, which sets the energy scale of the problem. 
Below we will give the temperature $k_B T$ in reduced units.
For example, for $a \sim
8\:$a.u. and $\epsilon \sim 10$ (typical values
of BMN solid solutions) in Eq.~(\ref{BV_Bsite}), 
$1350$ C corresponds to $k_B T=0.41$ in the simulation.

\begin{figure}[pt]  
\epsfig{file=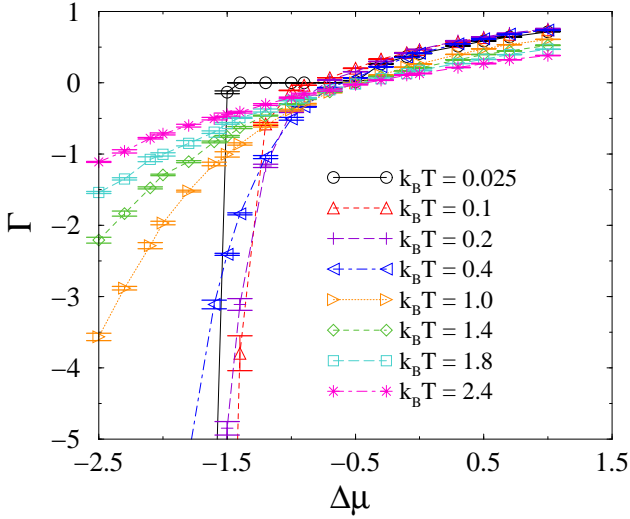} ,angle=-90, scale=.4 } 
\caption{Rocksalt growth rate vs. chemical potential for a [001] slab.}
\label{fig-growrate-NaCl}
\end{figure}

\begin{figure}[pt]  
\epsfig{file=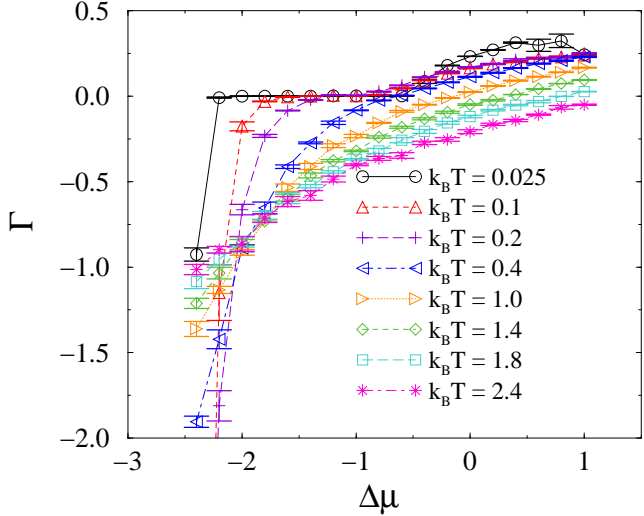},angle=-90,scale=.4}  
\caption{BMN growth rate  vs. chemical potential for a [-1-11] slab.}
\label{fig-growrate-BMN}
\end{figure}

\begin{center}
\begin{figure}[pt]  
\begin{tabular}{cc}

\epsfig{file=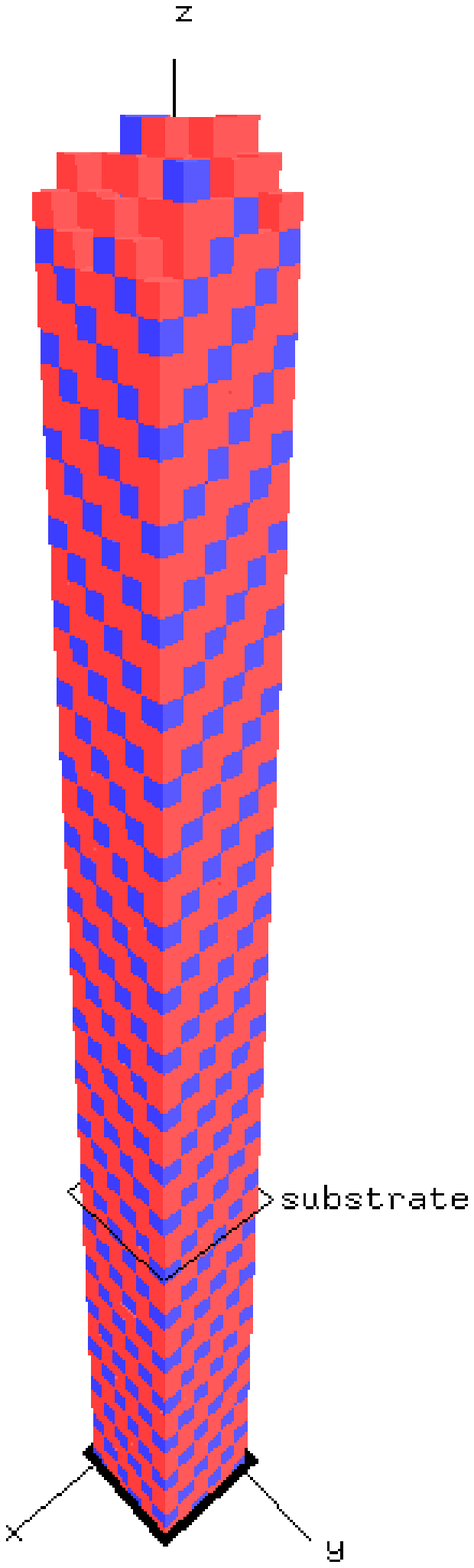} ,angle=0, scale=.35} &
\epsfig{file=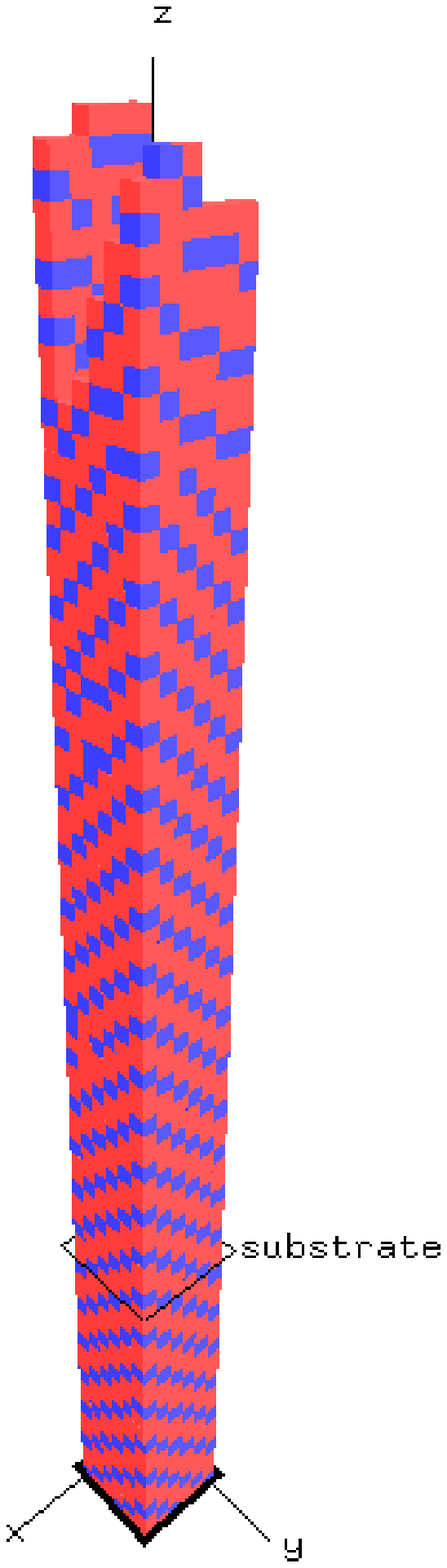} ,angle=0, scale=.35} \\
(a) & (b)\\
\end{tabular}

\caption {\label{visual_bmn2} \small Visualizations of 
grown BMN crystals. Shown are $6\times 6$ supercells with:
(a) growth direction along [001], $k_BT=0.1$ and $\Delta \mu=-1.0$;
(b) growth direction along [111], $k_BT=0.1$ and $\Delta \mu=-1.1$.
}
\end{figure}
\end{center}

As an overview, Figs.~\ref{fig-growrate-NaCl} and
\ref{fig-growrate-BMN} present a comparison of simulations of the
simple III$_{1/2}$V$_{1/2}$ rocksalt alloy and a II$_{1/3}$V$_{2/3}$ heterovalent alloy
such as BMN. 
(All substrates in our simulations have neutral
surface layers.)
We measure the growth rate of the crystal based on the 
KMC dynamics. 
If $N_G$ adatoms are gained in
$m$ MC steps (each defined as one attempt at the procedure outlined in
Section~\ref{sec:KMCnew}), the 
growth rate is defined as
\begin{equation}
\Gamma=\frac{N_G}{w_a\;\sum_{i=1}^m \Delta t_{\rm real}(i)}.
\label{eq:growthrate}
\end{equation}
Note that as defined the growth rate $\Gamma$ is renormalized by the absorption rate.
The growth rate is plotted as a function of
the chemical potential for a range of temperatures.
The rocksalt structure has
layers of positive and negative charges alternating along the [111]
direction.  It typifies the crystal ordering of a wide variety of
materials, including some of the perovskite alloys.  Heterovalent
binaries, described by II$_{1/2}$VI$_{1/2}$ ($q_B = \pm 2$) or
III$_{1/2}$V$_{1/2}$ ($q_B = \pm 1$), exhibit rocksalt B-site
chemical order.  By contrast, in the II$_{1/3}$V$_{2/3}$ heterovalent
binary BMN the equilibrium state shows [111]$_{1:2}$ ordering of two
layers of metal group V($q_B = +1$) alternating with one layer of the
group II($q_B = -2$) atom.  Both the rocksalt and BMN simulations
were initialized with a 20-layer  thick slab, with perfect [111]$_{1:1}$
 and [111]$_{1:2}$ ordering, respectively. The rocksalt simulation
used a 2-D $12 \times 12$ supercell, while the BMN simulations were done 
mostly with $6
\times 6 $ supercells, although some simulations with 
$12 \times 12$ and $15 \times 15$ were carried out to verify that 
the finite-size 
effects were small.  The rocksalt structure simulations ran
for $1,000L^2$ MC steps, up to a maximum thickness of 100 layers.  For BMN,
$10,000L^2$ MC steps were used, because for a given temperature and $\Delta
\mu$ growth was significantly slower.  
In Fig.~\ref{visual_bmn2}, we show visualizations of the grown 
BMN crystals to illustrate the simulation environment and the 1:2 order
at low temperatures with slow growth. 

The two sets of curves in Fig.\ref{fig-growrate-NaCl} 
and \ref{fig-growrate-BMN} are qualitatively
similar.  
What is not evident from the figures, however, is the degree of order
in each simulation. For a given temperature, as $\Delta \mu$
increases, the adsorbtion rate in Eq.~(\ref{eq:rateabs}) increases, and
adatoms are more likely to stick.  For fixed $\Delta \mu$, as $k_B T$
decreases, the adsorption rate will increase, but more importantly,
the ``selectiveness'' of evaporation will increase. A lower $k_B T$ will,
in effect, increase the energy differences between competing
configurations. The direct result, as growth is concerned, will be
that adatoms will increasingly prefer to have more 
instead of less neighbors with correct charge ordering
(layer-by-layer growth vs.\ rough growth), and 
adatoms with the same charge will seem more repulsive.  For very high $\Delta \mu$, adatoms
will stick anywhere, no matter what the location or ionic adversity
is, and the growth rate will be high. Alternatively, if the
temperature becomes too high, the crystal will melt, preferring the
liquid phase, and result in negative growth.  


To examine the degree of ordering,
we computed the charge-charge correlation
 function.
The Fourier transform of this correlation
 function, which we will denote by $\eta(\bm k)$, 
gives the structure factor:
\begin{equation}
\eta({\bm k}) = \alpha \sum_{{\bm ll'}} q_{\bm l}q_{\bm l+l'} 
\exp(-i{\bm k} \cdot {\bm l'})
\label{eta} 
\end{equation}
where $\alpha$ is the normalization factor, and  ${\bm k}$ is the  wave vector 
in the Brillouin zone of the unit cell.
The magnitude of $\eta(\bm k)$ characterizes 
the B-site order,
e.g., a large 
value of $\eta$ at ${\bm k } = \frac{2\pi}{a}\,(\frac{1}{2},\frac{1}{2},\frac{1}{2})$ 
 indicates a strong [111]$_{1:1}$ order
 while one at ${\bm k } = \frac{2\pi}{a}\,(\frac{1}{3},
\frac{1}{3},\frac{1}{3})$ indicates a strong [111]$_{1:2}$ order.

\begin{figure}[pt]  
\epsfig{file=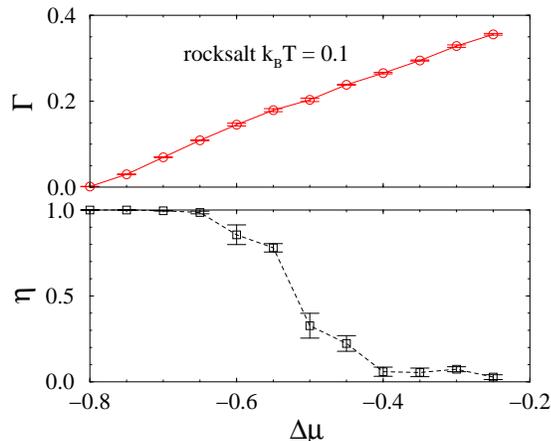} ,angle=-90, scale=.30} 
\caption{Rocksalt growth rate $\Gamma$ of Eq.~(\ref{eq:growthrate})
(top panel) and 1:1 order
         parameter 
$\eta({\bm k}=\frac{2\pi}{a}\,(\frac{1}{2},\frac{1}{2},\frac{1}{2}))$
(bottom panel) vs.~chemical potential. The temperature is
$k_B T = 0.1$ and the growth direction is [001]. 
A 12 $\times $ 12 supercell is used, with
         1000 MC time steps.
}
\label{gr-eta_nacl}
\end{figure}

\begin{figure}[pt]  
\epsfig{file=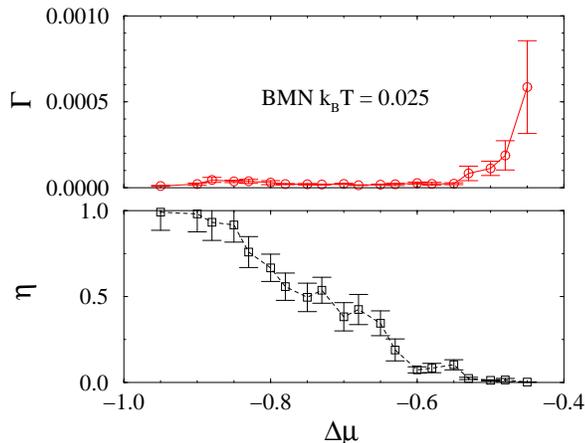},angle =-90,scale = .30} 
\caption{BMN growth rate $\Gamma$ of Eq.~(\ref{eq:growthrate}) (top panel) and 1:2 order parameter
$\eta({\bm k}=\frac{2\pi}{a}\,(\frac{1}{3},\frac{1}{3},\frac{1}{3}))$
         (bottom panel)  vs.~chemical potential. The temperature is
$k_B T = 0.025$ and the growth substrate direction is [$\bar{1}\bar{1}1$].
A 6 $\times $ 6 supercell is used, with
         300,000 MC time steps.}
\label{gr-eta_bmn0025}
\end{figure}

\begin{figure}[pt]  
\epsfig{file=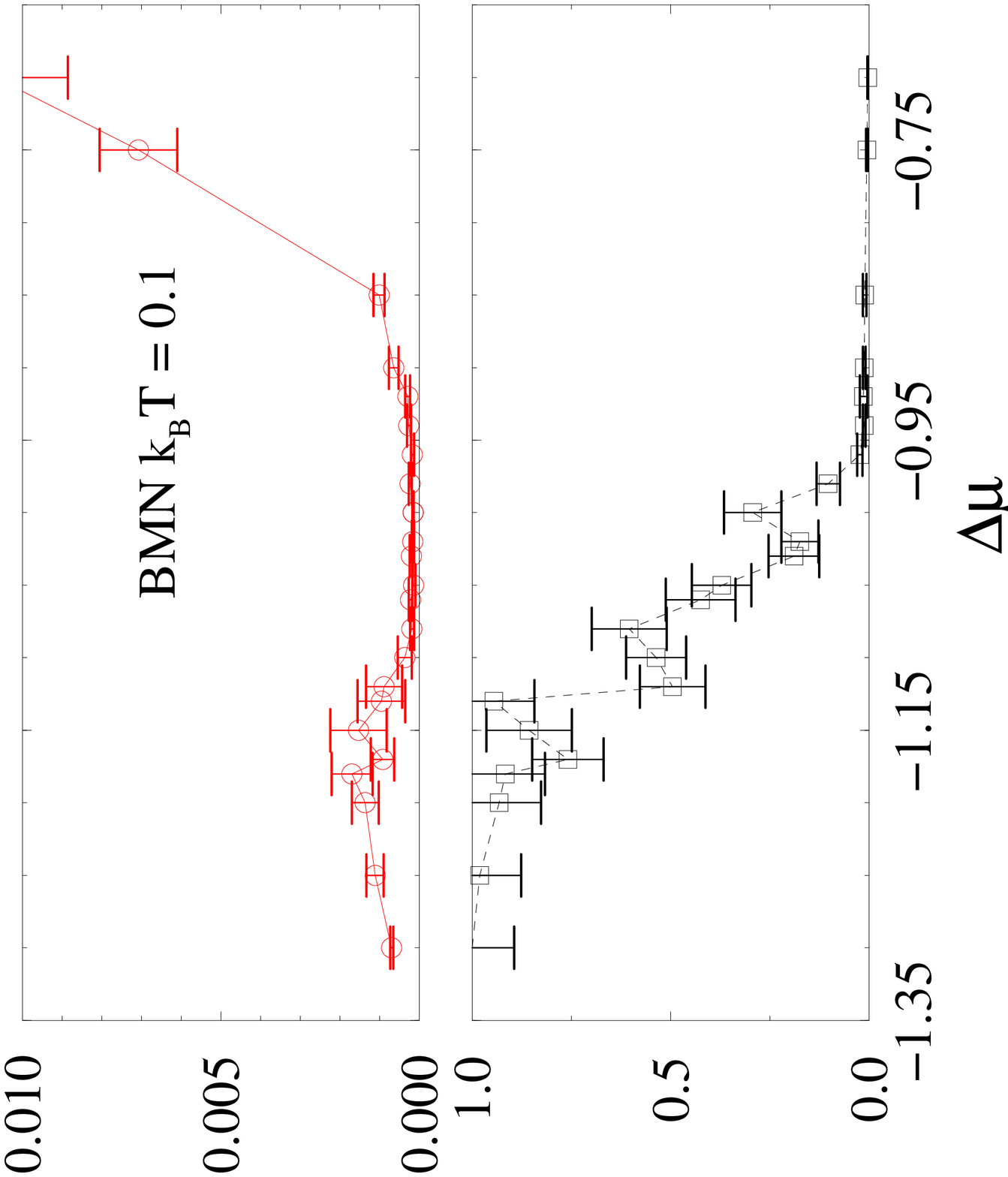} ,angle=-90, scale=.300} 
\caption{ BMN growth rate and 1:2 order parameter vs.chemical potential. The temperature is $k_BT$ = 0.1. Other parameters are the same as in Fig.~\ref{gr-eta_bmn0025}.}
\label{gr-eta_bmn01}
\end{figure}

\begin{figure}[pt]  
\epsfig{file=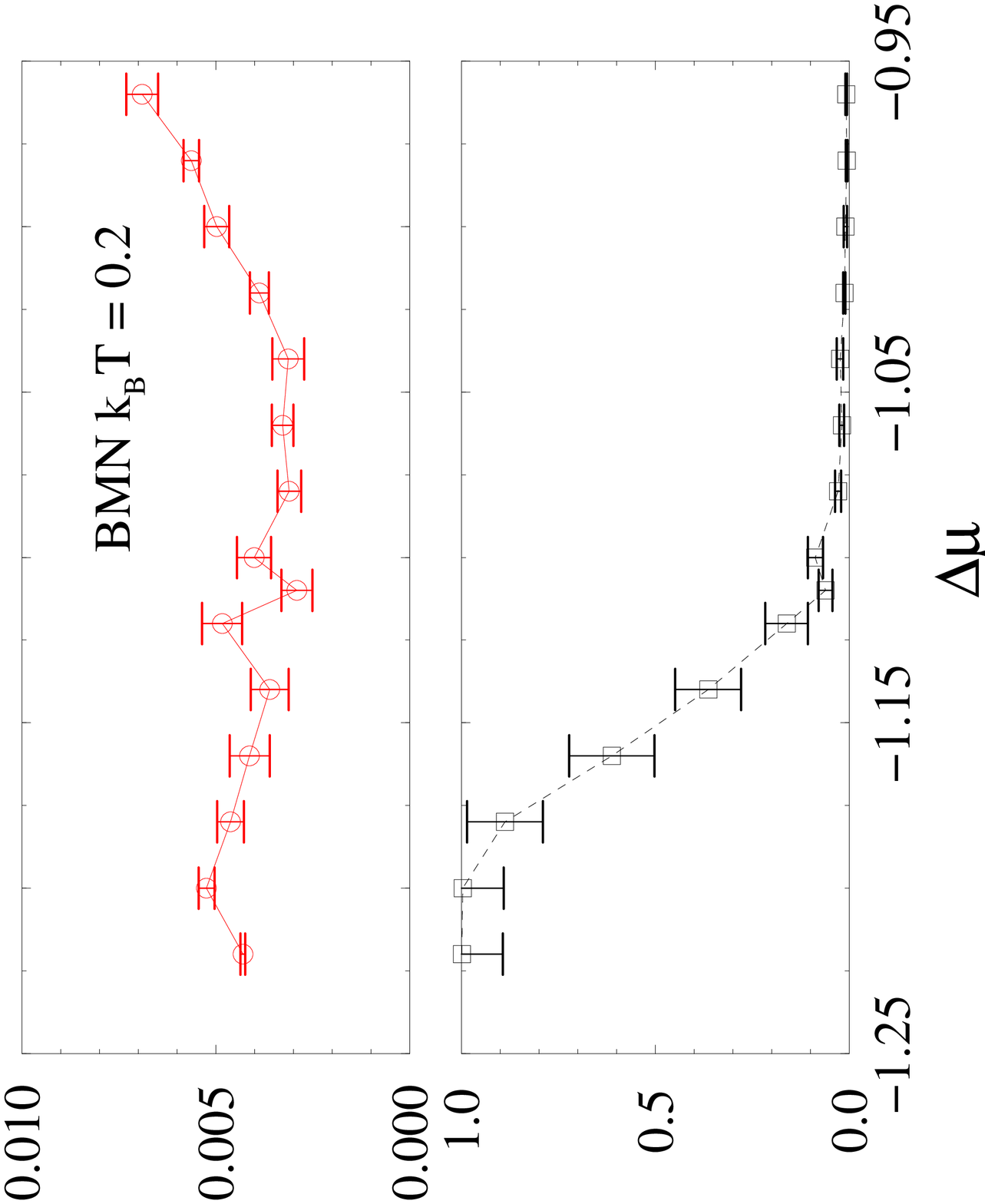},angle =-90,scale = .3}
\caption{BMN growth rate and 1:2 order parameter vs.chemical potential. The temperature is $k_BT$ = 0.2. Other parameters are the same as in Fig.~\ref{gr-eta_bmn0025} .}
\label{gr-eta_bmn02}
\end{figure}

The growth rate and the charge-charge structure factor
in Eq.~(\ref{eta}) are
plotted in Figs.~\ref{gr-eta_nacl}-\ref{gr-eta_bmn02}. 
In each
figure, the displayed range of $\Delta \mu$ was chosen to coincide with the range
where the order parameter decreases from nearly unity (perfect order)
to essentially zero (disorder). As $\Delta \mu$ increases the adsorption
rate increases, but the growth is disordered and there is greater
surface roughness.
Indeed there is only a limited range where ordered growth occurs.
The grown crystal structures are consistent with the observed
ground state configuration of rocksalt (Fig.~\ref{gr-eta_nacl}) and 
BMN (Fig.'s~\ref{gr-eta_bmn0025}-\ref{gr-eta_bmn02}).
The most striking difference between the growth behaviors of
rocksalt and BMN is the enormous reduction
of the growth rate of BMN compared to that of the rocksalt structure.
Moreover, for rocksalt the growth rate increases linearly as a function
of $\Delta \mu$ in the region where the order parameter $\eta$ is rapidly
decreasing. By contrast, the BMN growth rate is relatively constant
in this region. As $\Delta \mu$ increases beyond this region, there is a
sudden onset of much larger growth rates, but the resulting crystals are disordered. The growth rate of BMN increases as the temperature is increased
(Figs.~\ref{gr-eta_bmn0025}-\ref{gr-eta_bmn02}).

We next attempted to model the growth of BMN-BZ
(1-$x$)~(Mg$_{1/3}$Nb$_{2/3}$)~+~$x$~Zr solid solutions.  In the
electrostatic Hamiltonian in Eq.~(\ref{BV_Bsite}), tetravalent Zr
corresponds to a neutral charge $q_{\bm l} = 0$, so sites occupied by Zr have zero
interaction energy. As in the simulations of pure BMN systems, 
the chemical composition determines the probabilities with which
different charge species are adsorbed at the surface.
In the initial substrate, tetravalent ions with the corresponding
concentration were incorporated, using 
random mixing (next section). 
With a 1:2-ordered substrate, 
we studied concentrations $x \sim 10\%$, with temperatures of 
$k_BT \sim 0.1$ to $0.2$, 
and varying the chemical potential $\Delta\mu \sim-1.0$ to $-0.5$. 
Very little incorporation of the tetravalent ions occurred.
We found 
similar results with an initially 1:1-ordered substrate (random-site model; see below), where a wider range of
$x$ was explored. Again the order of the substrate
was not sufficient to
induce the incorporation of tetravalent ions 
in the growth phase. Instead the system seemed to favor
evaporating the adsorbed tetravalent ions more than the charged particles, to
grow pure BMN. 

\subsection{Energy Calculations}

To further study the inability to incorporate tetravalent ions at low temperatures, we examined
the total energy per particle $\varepsilon_N$ of fixed slab
configurations of B-site order. A phase separated model,
in which all the tetravalent adatoms were situated in the outermost
surface layers, was compared with various structural models that incorporated
tetravalent ions.  In each model, the calculations were performed for two
different configurational B-site orderings of the +2 and +5 ions
($q_{\bm l} = -2 , +1$, respectively). These configurations were the
1:1 and 1:2 layering along [111] directions, i.e.,
[111]$_{1:1}$ and [111]$_{1:2}$ order, repectively. 

The [111]$_{1:2}$
ordering corresponds to the $x=0$ order of BMN, with a layer of
$q_{\bm l} = -2$ alternating with two layers of charge $q_{\bm l} =
+1$ along the [111] direction.  We chose the [111]$_{1:1}$ ordering to
correspond to the random-site model \cite{ad}, which is observed in
the BMN-BZ equilibrium simulations for $x > 0.05$.
\cite{Bellaiche-Vander-98,WuKrak}
In the random-site model there are [111]
layers of $q_{\bm l} = +1$ alternating with a mixed layer of charges
$q_{\bm l} = -2 , +1, 0$.  The random-site model is meant to represent
the presence of short-range B-site order from experimental observations.
No long-range ordering has been observed. Nevertheless
in our simple model here we will fix the ordered $q_{\bm l} = +1$ layers  
and choose
the mixed layers to be a random mixture
of $(-2)_{\frac{2}{3}(1-x)}(+1)_{\frac{1}{3}(1-4x)}(0)_{2x}$. 

\begin{figure}[pt]  
\epsfig{file=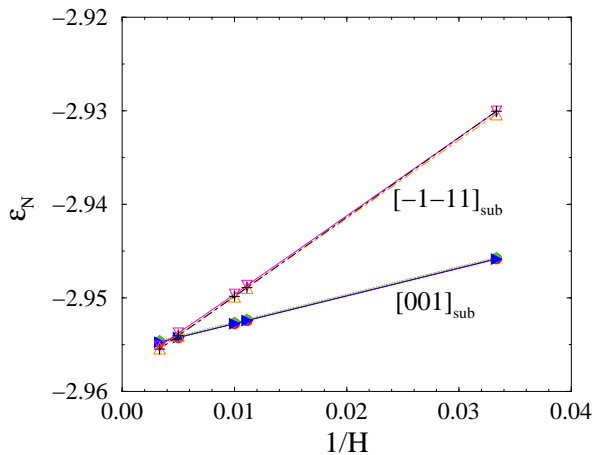} ,angle=-90, scale=.35}
\caption{Total energy per particle for B-site
         [111]$_{1:2}$ ordering as a function of slab 
         thickness $1/H$ and slab crystallographic orientation. 
         Each set has three barely distinguishable curves, 
         corresponding to three lattice sizes: $12 \times 12$, 
        $15 \times 15$, and  $18 \times 18 $.}
\label{size_effect} 
\end{figure}

We first examine finite-size effects in Fig.~\ref{size_effect}, which
plots $\varepsilon_N$ as a function of slab thickness for various 2-D
supercells containing no tetravalent ions, for [111]$_{1:2}$ ordering.
Results for [001] and 
[$\bar{1}\bar{1}1$]
slabs are shown, both of which correspond to neutral surface
layers.  
As $H \rightarrow \infty$, 
$\varepsilon_N \sim \varepsilon^B_N +{\rm const.}/H$ 
as expected, where the constant 
$\varepsilon^B_N$ represents the average bulk value and 
$H$ is the slab thickness.


\begin{figure}[pt]  
\epsfig{file=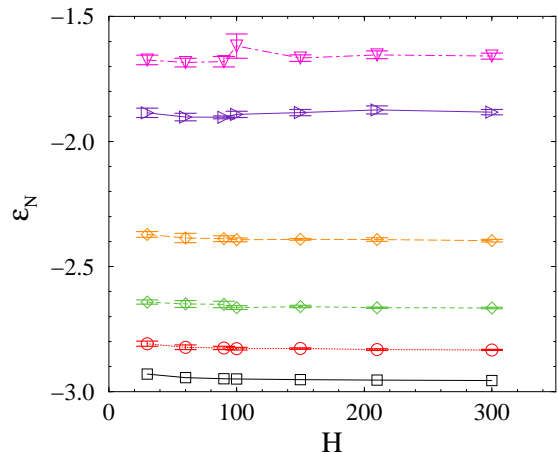} ,angle=-90, scale=.35} 
\caption{Total energy per particle for [111]$_{1:2}$ ordering, as a function of slab thickness
         $H$ and (randomly mixed) tetravalent concentration $x$. E/N increases, as x increases from 0\% to 2\%, 5\%, 10\%, 15\%, and 25\%. 
        A [$\bar 1 \bar 11$] slab with a $15
         \times 15$ supercell was used.}
\label{wn-percentage}
\end{figure}

\begin{figure}[pt]  
\epsfig{file=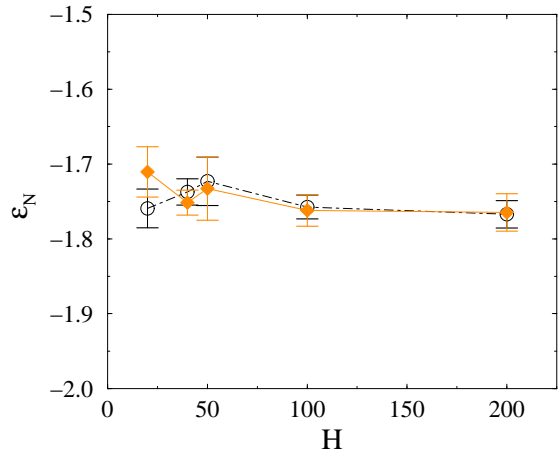} ,angle=-90, scale=.35}
\caption{Total energy per particle vs. slab thickness.
 Results are shown for [111]$_{1:1}$ ordering with (dashed line) and
         without (solid line) 10\% randomly mixed tetravalent ions.
      A [001] slab with a 12 $\times$ 12 supercell was used.}
\label{orders_11}
\end{figure}

Size effects with incorporated tetravalent ions are studied next. 
Fig.~\ref{wn-percentage} plots $\varepsilon_N$ for [111]$_{1:2}$ ordering
as a function of slab thickness for various concentrations of randomly
mixed tetravalent ions, using a [$\bar 1 \bar 11$] slab and 15 $\times$ 15 supercell.  These
calculations are for a random distribution of +0 (tetravalent) ions replacing -2 or
+1 ions in an otherwise perfectly ordered [111]$_{1:2}$ slab at each 
thickness $H$. Within statistical error bars, the asymptotic
$H$-dependence is similar to that without tetravalent ions in
Fig.~\ref{size_effect}.
Fig.~\ref{orders_11} plots $\varepsilon_N$
for [111]$_{1:1}$ ordering with and without randomly mixed 10\% tetravalent ions.
As seen in Fig.~\ref{wn-percentage}, $\varepsilon_N$ rapidly increases
with increasing $x$.
This is consistent with the inability to
incorporate tetravalent ions in the growth simulations on [111]$_{1:2}$
ordered slabs. 
By contrast 
$\varepsilon_N$ for [111]$_{1:1}$ ordering is essentially independent of $x$ within statistical
error, as shown in Fig.~\ref{orders_11} 

\begin{figure}[pt]  
\epsfig{file=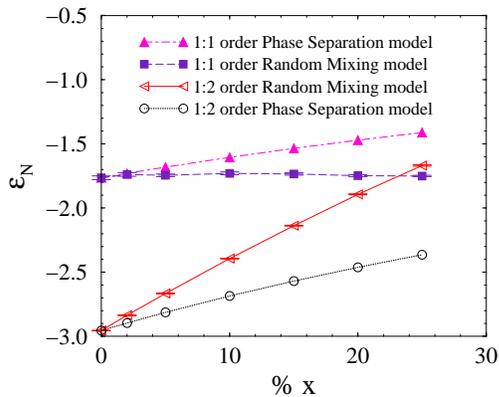} ,angle=-90, scale=.30} 

\caption {Total energy per particle vs. tetravalent concentration $x$
          for random-mixing and phase-separation models. Results are
          shown for [111]$_{1:1}$ and [111]$_{1:2}$ ordered $12 \times
          12 $ [001] slabs ($H = 200$).}
\label{ps-rm-1-2-order}
\end{figure}

Fig.~\ref{ps-rm-1-2-order} plots $\varepsilon_N$ as a function of
tetravalent concentration $x$ for random-mixing and phase-separation
models, showing results for [111]$_{1:1}$ and [111]$_{1:2}$ ordered
$12 \times 12 $ [001] slabs ($H = 200$). For the phase-separation
model, the total number of ions includes the outermost layers of
tetravalent ions. 
At $x = 0$ the 1:2 ordered
crystal has a lower energy than the 1:1 ordered crystal, which is
consistent with our results from the growth simulation 
and with 
the observed ground state configuration of pure
BMN. 
For random-mixing, $\varepsilon_N$ increases
linearly with $x$ for [111]$_{1:2}$ ordering while it is essentially
independent of $x$ for [111]$_{1:1}$ ordering. In the phase-separation
model, $\varepsilon_N$ increases linearly for both orderings. These
results show that phase separation is favored for the [111]$_{1:2}$
ordering, while random mixing is favored by [111]$_{1:1}$ ordering. 

\begin{center}
\begin{figure}[pt]  
\begin{tabular}{cc}

\epsfig{file=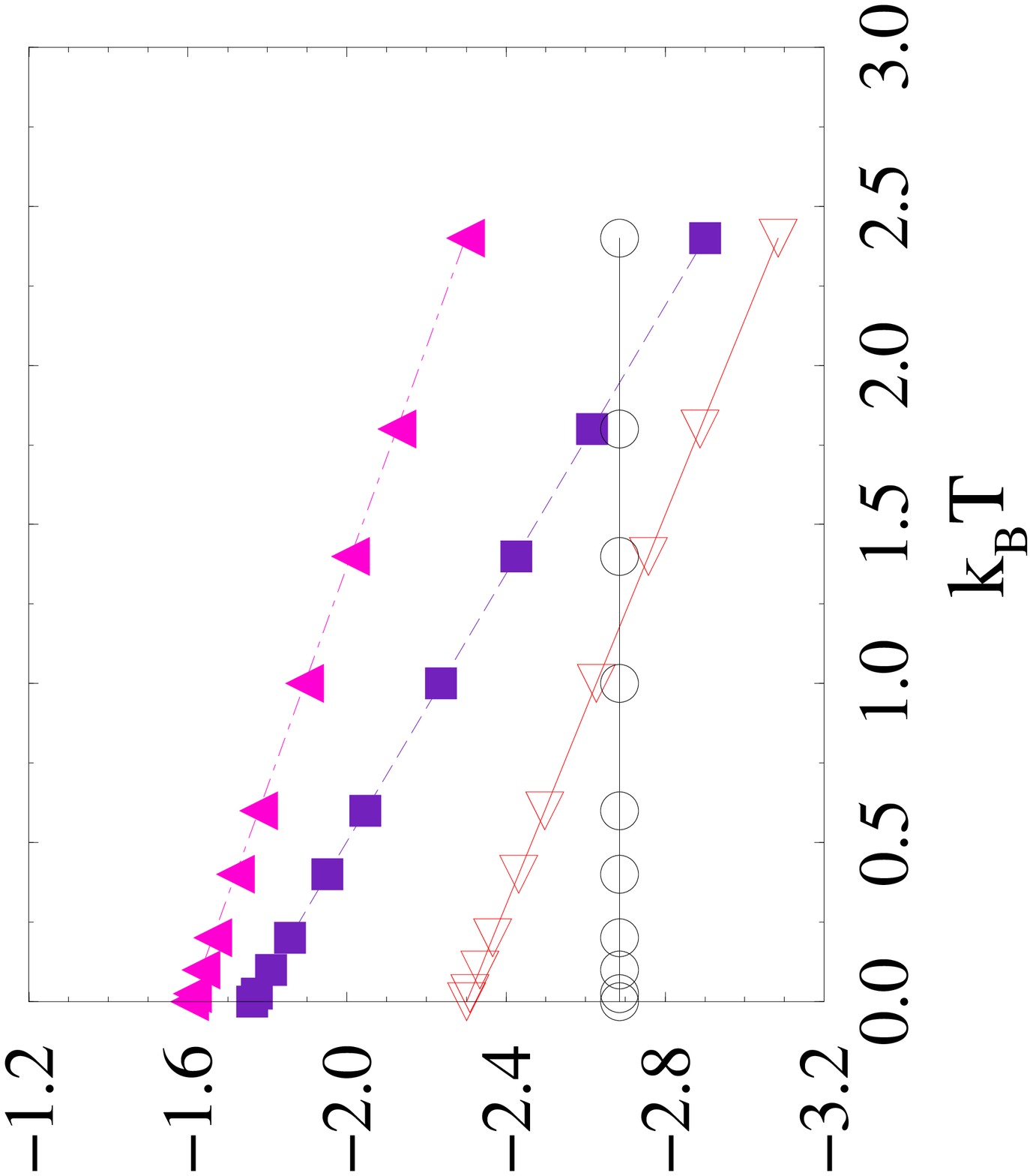} ,angle=-90, scale=.20} &
\epsfig{file=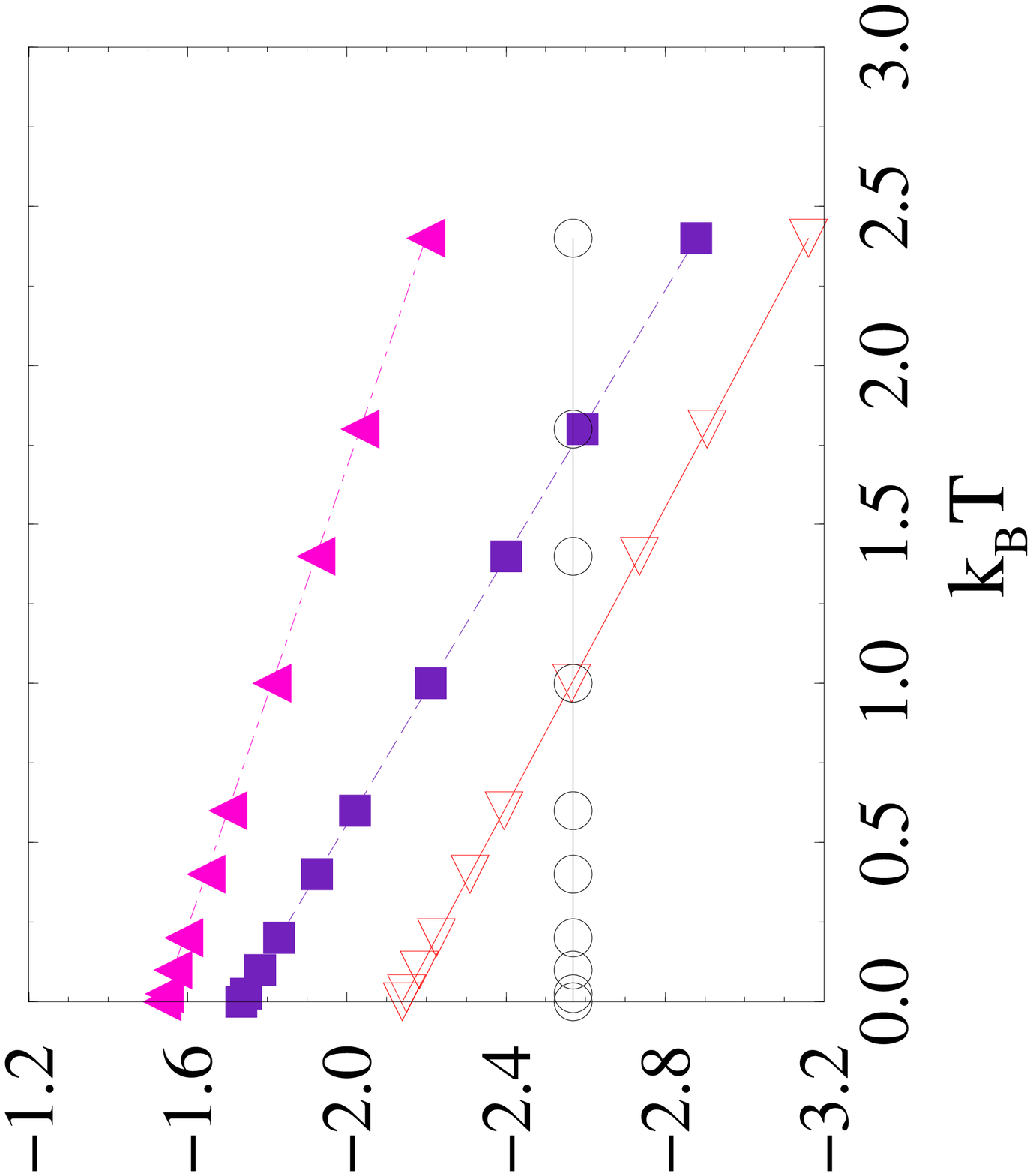} ,angle=-90, scale=.20} \\
\\
(a) & (b)\\
\epsfig{file=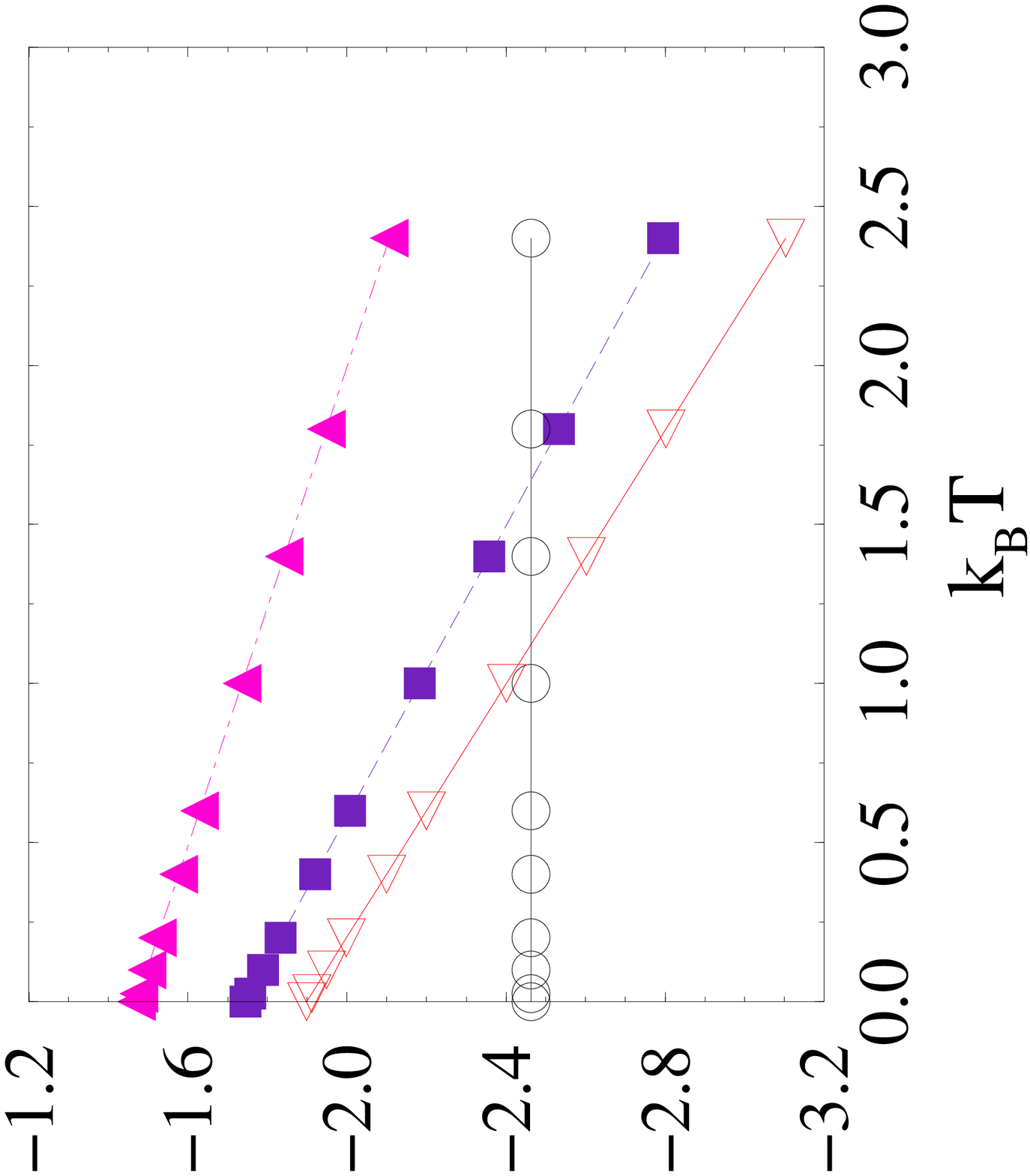} ,angle=-90, scale=.20} &
\epsfig{file=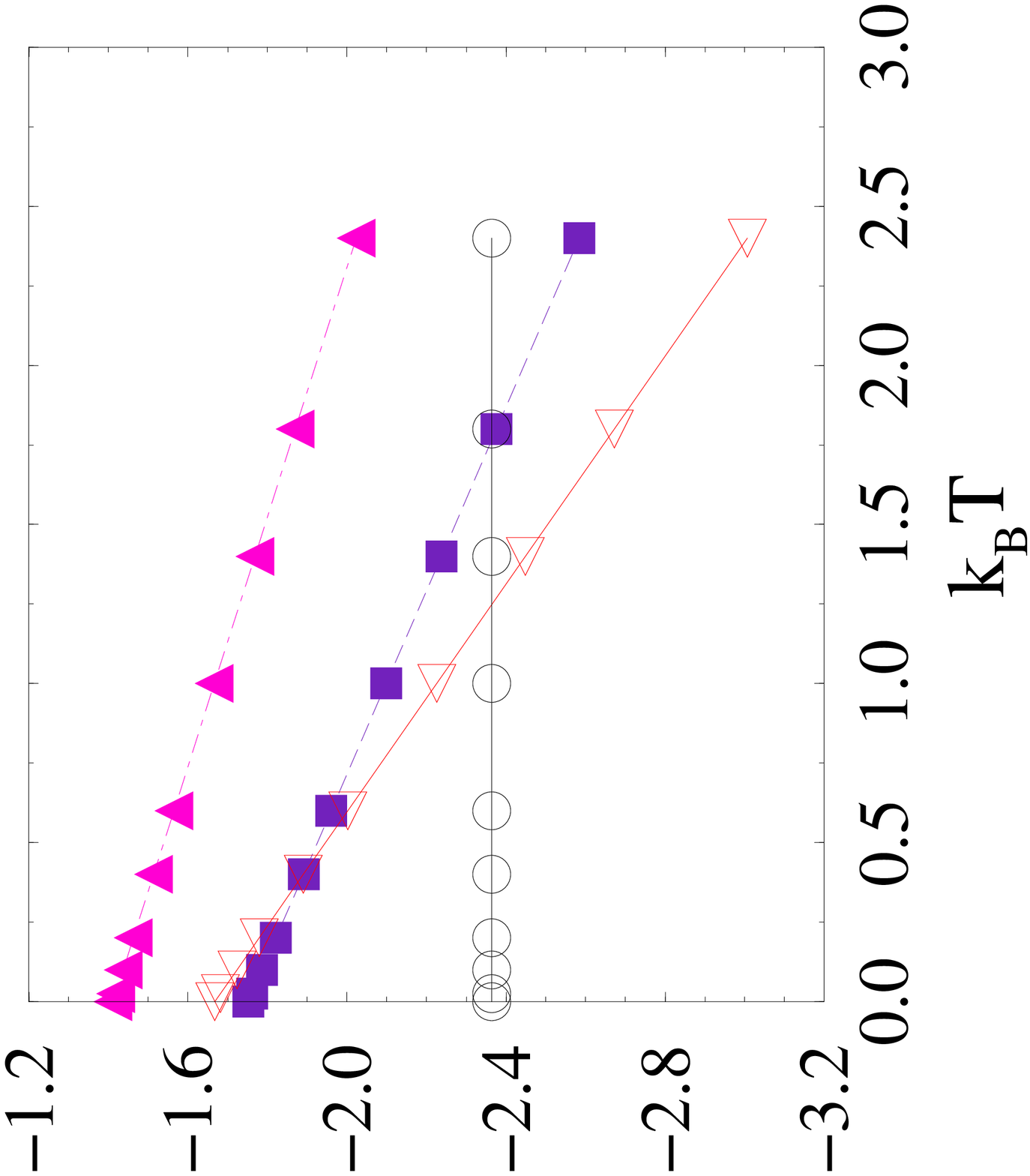} ,angle=-90, scale=.20}\\
\\
(c) & (d) 
\end{tabular}

\caption {\label{free-energy} \small Free energy of BMN crystal for
(a) 10\% (b) 15\%  (c) 20\%  (d) 25\% tetravalent concentrations. 
Symbols have the same meaning as in Fig.~\ref{ps-rm-1-2-order}.}
\end{figure}
\end{center}

 Fig. \ref{ps-rm-1-2-order} illustrates why 
the growth simulations failed to incorporate tetravalent ions at low temperature. In the
electrostatic model, the 1:2 ordered state is the ground state and is
optimally ordered. The potential energy between any charge and all other charges in
the system is negative. For example, with a $18\times18$ slab this potential 
energy is $\sim -5.92$ for a $-2$ charge and $\sim -1.48$ for a $+1$ charge.  
Thus, replacing a charge (either $\,-2$ or $+1\,$) by a neutral tetravalent ion in this
state raises the total energy of the system, while a phase-separated
configuration in which the tetravalent ion is placed away from the the ordered slab
keeps the total energy unchanged. 
To examine this more closely, we calculated the free-energy ($F =
\varepsilon_N - T S$), where $S$ is the mixing entropy due to the
incorporated tetravalent ions.  Fig.~\ref{free-energy} plots the free-energy
as a function of temperature for four concentrations of tetravalent ions.  The
free energy of the phase-separated 1:2 ordered slabs is constant in
our model, because it is perfectly ordered and has vanishing
entropy.  The free energy of the phase-separated 1:1 ordered slabs
decreases with increasing temperature, despite the perfectly
ordered outmost layers of tetravalent ions, due to the mixing entropy of the
random layers with $-2$, $+1$, and $0$ charges.  In all cases in
Fig.~\ref{free-energy}, the phase-separated 1:2 ordered slabs have the
lowest free energy at low temperatures, where ordered crystal growth 
occurs in our simulations, but at temperatures between $k_B T \sim 1 - 2 $
the 1:2 ordered and 
the 1:1 ordered random mixing models start to be favored.

\section{Discussion}

There are striking differences between the growth behavior of the III$_{1/2}$V$_{1/2}$ rocksalt ordered structure
and the II$_{1/3}$V$_{2/3}$ BMN structure. 
The ordered rocksalt structure forms over a wide range of
$\Delta \mu$ (absorption rates) as shown in Fig. 4.
By contrast, ordering of the 1:2 structure in
BMN type crystals is more difficult to achieve experimentally. \cite{Davies-Tong-1997,Akbas-Davies-1998}
%
%
When these materials are initially synthesized, they crystallize in a disordered structure. 
With extended annealing the 1:2 structure is approached. \cite{Davies-Tong-1997} 
As discussed by Davies {\it et al.} \cite{Davies-Tong-1997}, the initial synthesis and processing are 
controlled by irreversible kinetic processes rather than by thermodynamic factors, and a more correct
description of the formation of the 1:2 ordered structures is in terms of the nucleation and growth of small
ordered domains with increasing annealing time and temperature.
Eventually large ($>$100 nm) 1:2 ordered domains are
observed.\cite{Davies-Tong-1997,Akbas-Davies-1998}  The need for long annealing times
is consistent with our simulations.  Figs. 5-7 show that the range of
$\Delta \mu$ where ordered 1:2 growth occurs narrows as
the temperature increases from $k_B T$ = 0.025 to 0.2.  In this range,
the growth rate is approximately constant as a function of $\Delta
\mu$.  Moreover, when ordered crystal
growth occurs, the BMN growth rate is much smaller than that of the rocksalt structure
at the same temperature.  Highly ordered growth was possible in the BMN simulations
but required low temperatures and a delicate balance with the
chemical potential. Neither of these requirements is likely to be met
under experimental synthesis conditions.  At temperatures
corresponding to the actual sintering temperature of BMN ($k_B T \sim
0.5$), large growth rates can be achieved, as shown in Fig. 2, but the
growth is highly disordered.  The long annealing times 
allow the slow formation of the 1:2 ordered regions. In our KMC
simulations, diffusion processes are excluded so there can be no
annealing.  We also note that the growth rate was
sensitive to the slab orientation. For example, we found that growth
rate along [$\bar 1 \bar 11$] direction was almost an order of
magnitude larger than that along [$001$].

Our results are also qualitatively consistent with the long
experimental history of failed attempts to coarsen the 1:1 ordered
nanoscale domains in PMN type crystals.  Prior to the 
experiments of Akbas and Davies \cite{Akbas-Davies-1998}, 
the 1:1 ordered regions were apparently limited to
nanoscale size and represented only a small volume fraction of the
crystal. The space-charge model, which was invoked to explain this
behavior, hypothesized that the 1:1 ordered regions arose from a
rocksalt ordering of the -2 and +1 B-site charges, implying
charge-imbalanced 1:1 domains. The apparently limited size of these
domains could be explained by the rapidly increasing energy of larger
domains due to Coulomb repulsion. With careful annealling at
much higher temperatures than had previously been tried, however,
some fully 1:1 ordered crystals were synthesized \cite{Akbas-Davies-1998}. Our
calculations show that long-range ionic interactions favor the growth
of disordered crystals, and ordering occurs only after annealing.
Moreover, ionic interactions appear to favor the 1:2 ordering.  
However, entropic
contributions to the free energy and short-range covalent interactions
tend to favor 1:1 ordering. Covalent bonding is negligible for Ba ions
but very important for Pb ions. Thus there is a delicate competition
between 1:2 and 1:1 ordering for doping with small concentrations of the
tetravalent ions in (1-x)BMN-xBZ and (1-x)PMN-xPT.  In (1-x)BMN-xBZ,
there is a crossover from 1:2 to 1:1 ordering as $x$ increases to
about 5\%. While in (1-x)PMN-xPT, the stronger short-range covalent
bonding of Pb favors 1:1 ordering at all concentrations.

For pure systems, our minimal paradigm for growth simulations 
captures the differences in growth rate and ordering between rocksalt-type and BMN-type
crystal growth.  This indicates that the simple ionic model is a
reasonable starting point for describing the growth of perovskite 
solid solutions.  
More direct and quantitative comparisons with experiment 
will require additional ingredients such as
short-range interactions and the inclusion of diffusive processes.

For systems with tetravalent ions, 
our results show that the ground state is a phase-separated
state of tetravalent ions and 1:2 ordered BMN 
over a wide range of tetravalent compositions.
On the other hand, equilibrium simulations of the ionic 
model \cite{Bellaiche-Vander-98,WuKrak} suggest that
for $x > 0.05$ the 1:1 ordering is preferred, with no phase separation. 
Several factors distinguish these calculations, which likely have to do with the
apparent contradiction in their observations. 
The first is the difference in the nature of the 
simulations. 
In our growth simulation, tetravalent ions are allowed to evaporate from the 
crystal, which facilitates phase separation. The equilibrium calculations
were done in the canonical ensemble with the tetravalent ions 
mixed in, where it is more difficult to detect phase separation without 
large simulation cell sizes. 
Our simulations
were at lower temperatures where ordered growth could be induced by tuning the
chemical potential $\Delta \mu$ (absorption rate). 
At these temperatures the system is essentially in the ground state,
as Fig.~\ref{visual_bmn2} shows.
Incorporation of tetravalent ions could be induced at
larger $\Delta \mu$, which is expected as adsorption dominates
evaporation, but in this case random growth occurs.
Secondly, since our [111]$_{1:1}$ structure is an artificial model of {\em random mixing\/}
of $-2$, $+1$, and neutral charges in one layer and perfectly ordered 
$+1$ in another, its energy must be higher than the actual 1:1 structure achieved
in the equilibrium simulations.
This means that the actual cross-over of the random-mixing 
[111]$_{1:1}$ structure will occur at lower 
temperatures.
Indeed, the $k_BT \sim 0.25$ equilibrium calculations show [111]$_{1:1}$ ordering
for concentrations $x$ greater than about 0.05.
Thus the absence of
phase separation in the equilibrium calculations might be due to a lower free-energy
than our estimate in Fig.~\ref{free-energy} from the artificial random-site structure.
Our results combined with the equilibrium calculations therefore
suggest the following picture of the equilibrium state of the ionic model.
In the ground state phase-separation takes place for $x>0$. 
Beyond some  $x$-dependent critical temperature tetravalent ions are incorporated, most likely in a structure that 
favors 1:1 order.

To determine if 
the new phase (phase-separation) at low temperatures that we have found 
is realistic for these alloys, the ionic model must be improved. 
One possibility is 
first-principles based $H_{\rm eff}$, which have shown great promise
in describing ferroelectrics and simple solid-solutions
\cite{Ferro2002}. Like the Ising model these $H_{\rm eff}$ project out
what are considered to be the most important ionic degrees of freedom.
In addition to the long-range Coulomb interaction, short-range interactions
are also included.
The $H_{\rm eff}$ parameters are fitted to the results of
a set of first-principles density-functional calculations, so there is
effectively no experimental input (except sometimes the average
crystal volume). The simplified form of $H_{\rm eff}$ for ferroelectrics
and ferroelectric alloys has permitted simulations of equilibrium
properties on thousands of atoms as a function of temperature and
applied external electric field. A main difficulty in applying these
in a growth simulation is computational cost, which has typically  
required fixed distributions of B-site ions even in 
equilibrium simulations of solid-solutions. 
In our kinetic Monte Carlo model, another possibly important factor that is not 
included is surface diffusion. Coupled with the solid-on-solid restriction,
the simulation is severely limited in its ability to ``heal'' disorder, and these approximations
may have contributed to low ordered growth rates and raised the critical temperature for phase separation.
Removal of these 
restrictions would improve the model and increase its applicability. 

\section{Summary}

The growth of the technologically important BMN type perovskite alloys 
was studied by kinetic Monte Carlo using an ionic model.
An enhanced KMC algorithm was formulated to  
treat long-range Coulomb interactions
efficiently. We found that this minimal paradigm was capable of 
describing ordering features of the growth of pure BMN and PMN type single crystals.
The largest growth rates were observed along the [$\bar 1 \bar11$]
direction, but best ordered growth rates are substantially less than those of
rocksalt. Highly ordered growth was possible, 
but required very low temperatures
and a delicate balance with the chemical potential.
For mixed systems such as BMN-BZ, we found that the $T=0$ ground state
of the model was one in which tetravalent ions phase separate from 
a 1:2 ordered pure system. As a result, little incorporation of tetravalent ions 
occurs in the growth process at low temperatures. 
At higher temperatures,
tetravalent ions can be incorporated, but the resulting crystals show 
no chemical ordering.
The tendency of the purely ionic model to favor phase separate was further studied
using free energy calculations determined from T = 0 total energy calculations and including a
mixing entropy. This indicated that,
if diffusive mechanisms were included, chemical orderings
consistent with those found in equilibrium studies could develop
at the higher
temperatures characteristic of realistic alloy synthesis.

\begin{acknowledgments} 

Support from ONR (N000149710049 and N000140110365), NSF (DMR-9734041), and 
Research Corporation are gratefully acknowledged. We would like to thank
C.~Tahan, T.J.~Walls, and P.~Larsen for enjoyable and productive 
collaborations and for their contributions in early stages 
of this work.

\end{acknowledgments}

\appendix

\section{Coulomb potential}

The 2-D Ewald potential, is given as the sum of three terms
\begin{equation}
v (\bm{l}'-\bm{l})=v_1 (\bm{l}'-\bm{l})+v_2 (\bm{l}'-\bm{l})
+v_s (\bm{l}'),
\end{equation}
where $v_1$ and $v_2$ are due to $\rho _1 ({\bm{r}})$ and $\rho _2
({\bm{r}})$, respectively in Eq. (\ref{rho1rho2}), and $v_s$ is the
correction for the interaction of the point charge $q_{\bm{l' }}$
with its own Gaussian density $q_{\bm{l' }} g(\bm{r-\tilde l'})$
in $\rho _2 ({\bm{r}})$.

To calculate $v_1 (\bm{l}'-\bm{l})$ we place, for consistency,
the ($\bm{R}$~$\ne$~0) $q_{\bm{l' }}$ images  at their
vertical projections onto the plane of the $q_{\bm{l }}$
sublattice. $v_1 (\bm{l}'-\bm{l})$  is then given by
\begin{eqnarray}
  v_1 ({\bm{l }}'{\bm{ - l }}) 
   &=& q_{\bm{l }} \sum\limits_{\bm{R}} {\frac{{{\mathop{\rm erfc}\nolimits} \left( {\sqrt \alpha  \left| {{\bm{l }}'{\bm{ - l  - R}}} \right|} \right)}}{{\left| {{\bm{l }}'{\bm{ - l  - R}}} \right|}}} \nonumber  \\ 
   &\ & - q_{\bm{l }} \sum\limits_{{\bm{R}} \ne 0} {\frac{{{\mathop{\rm erfc}\nolimits} \left( {\sqrt \alpha  \left| {{\bm{l }}'{\bm{ - \tilde l }}'{\bm{ - R}}} \right|} \right)}}{{\left| {{\bm{l }}'{\bm{ - \tilde l }}'{\bm{ - R}}} \right|}}}.
\label{V_1}
\end{eqnarray}
The mathematical form of this contribution is identical to its 3-D counterpart, except that
the sum is over 2-D rather than 3-D direct-lattice vectors $\bm{R}$.

The 2-D planewave expansion of
$\tilde \rho _2 ^{({\bm{l }},{\bm{l '}})} ({\bm{r}})$ in Eq.~(\ref{rhotilde})
is given by
\begin{eqnarray}
 \tilde \rho _2 ^{({\bm{l }},{\bm{l '}})} ({\bm{r}}) = q_{\bm{l }} \left( {\frac{\alpha }{{\pi A^2 }}} \right)^{1/2} e^{ - \alpha (z - l _z )^2 } \hfil
\nonumber \\ 
 \sum\limits_{{\bm{G}} \ne 0} {
e^{ - G^2 /4\alpha } \left[ {e^{ - i{\bm{G}}\cdot{\bm{l }}_{\alpha p} }  - e^{ - i{\bm{G}}.{\bm{l }}'_{\alpha p} } } \right]
e^{  i{\bm{G}} \cdot \bm{r}_p}
} ,
\end{eqnarray}
where we have used the fact that $\bm{\tilde l
'}_z$~=~$\bm{l}_z$.  Substituting into Eq.~(\ref{V_quadr_G}) and
using Eq.~(\ref{Green_G}), yields:
\begin{equation}
\begin{array}{l}
 v_2({\bm{l }}' \!-\! {\bm{l }}) = \sum\limits_{{\bm{G}} \ne 0} {
{\textstyle{\pi  \over {AG}}} \left[ {f(G)\!-\!f(-G)} \right] 
\left[ {e^{i{\bm{G}}\cdot({\bm{l }}'_p  \!-\! {\bm{l }}_p )}  \!-\! 1} \right] 
} ,
 \end{array}
\end{equation}
where
\begin{equation}
f(x) \equiv e^{x(l '_z - l _z )} {\rm{erfc}}\left( {{\textstyle{{2\alpha \left| {l '_z - l _z } \right| + x} \over {2\sqrt \alpha  }}}} \right) .
\end{equation}

Finally, the correction for the interaction of the point charge $q_{\bm{l' }}$ 
with its own Gaussian density is given by:
\begin{equation}
v_s ({\bm{l }}') = \frac{{{\rm{erf}}(\sqrt \alpha  \left| {{\bm{l }}' - {\bm{\tilde l }}'} \right|)}}{{\left| {{\bm{l }}' - {\bm{\tilde l }}'} \right|}}.
\end{equation}

As verified by direct calculation, the sum of these three terms in
independent of the parameter $\alpha$. For efficiency, 
$v (\bm{l}'-\bm{l})$ is stored as a look-up table.

\bibliography{bibdb}

\end{document}